\documentstyle[12pt]{article}
\input epsf
\newcommand{\be}{\begin{equation}}
\newcommand{\ee}{\end{equation}}
\newcommand{\beq}{\begin{eqnarray}}
\newcommand{\eeq}{\end{eqnarray}}
\def\ket #1{|#1\rangle}
\def\bra #1{\langle#1|}
\def\matel #1#2#3{\bra{#1}#2\ket{#3}}

\begin{document}

\thispagestyle{empty}
\rightline{UCSBTH-94-20}
\rightline{hep-th/9407008}

\vskip 2.5 cm

\begin{center}
{\large\bf A POSSIBLE RESOLUTION OF THE
BLACK HOLE INFORMATION PUZZLE} \break

\vskip 1.0 cm

{\bf Joseph Polchinski}

{\sl Institute for Theoretical Physics  \break
University of California  \break
Santa Barbara, CA 93106-4030} \break

and

{\bf Andrew Strominger}

{\sl Department of Physics \break
University of California \break
Santa Barbara, CA 93106-9530} \break

\end{center}
\vskip 1.0 cm

\begin{quote}
{\bf ABSTRACT:}
The problem of information loss is considered under the  assumption that
the process of black hole evaporation terminates in the decay of
the black hole interior into a baby universe. We show that
such theories can be decomposed into superselection sectors
labeled by eigenvalues of the third-quantized baby universe field
operator, and that scattering is unitary within each superselection
sector. This result relies crucially on the quantum-mechanical
variability of the decay time. It is further argued that the decay
rate in the black hole rest frame is necessarily proportional
to $e^{-S_{tot}}$, where $S_{tot}$ is the total entropy produced
during the evaporation process, entailing a
very long-lived remnant.

\end{quote}

\newpage

In the seventies, Hawking \cite{hawk} made the profound discovery that
quantum mechanical black holes evaporate. Hawking went on to claim
\cite{hawktwo} that black holes ultimately disappear, and take with
them most of the information contained in the initial state which
formed the black hole. This claim ignited a controversy which has
continued up to the present. Three main schools of thought have
emerged on this black hole information puzzle:

(I) Information is destroyed in quantum processes involving black holes.

(II) A very careful analysis will reveal that the information comes back out.

(III) The information is stored in an eternal or long-lived remnant.

\noindent In this paper we shall present a fourth alternative, which
might be described as

(IV) All of the above.

\noindent In our proposal information is lost in the sense that
arbitrarily precise knowledge of the local laws of physics is
insufficient to predict the outcome of gravitational collapse.
Additional coupling constants (relatives of the $\alpha$-parameters of
wormhole physics \cite{worm}) are required which can only be measured
by forming black holes and watching them evaporate. After a very large
number of experiments, these parameters can be determined to within a
finite accuracy. Less and less information is then lost in each
successive experiment, and asymptotically the outcome becomes
completely predictable. A key ingredient providing for the
self-consistency of our picture is a remnant which remains after the
black hole horizon has shrunk to zero (or Planckian) size. Compatability
with information bounds on remnant lifetimes \cite{cash} imply that these
remnants must be very long-lived. We indeed give a dynamical argument
that in the models
under consideration the decay time is proportional to $e^{S_{tot}}$,
where $S_{tot}$ is the total entropy in the Hawking radiation produced
during the evaporation process.

Our analysis assumes the qualitative features\footnote{An
attempt to construct a two-dimensional model incorporating these features
was made in \cite{as} .} of black hole
evaporation depicted
in figure~1. A sufficiently energetic incoming pulse of matter
collapses into a black hole. The apparent horizon subsequently
shrinks, as expected from the usual semiclassical reasoning.
Eventually the apparent horizon reaches zero size. The spatial
geometry (for an appropriate slicing) then contains an exterior,
asymptotically flat region connected to the black hole interior by an
umbilical cord of Planckian dimensions. The umbilical cord breaks with
an amplitude proportional to $g_s$ per unit proper time along its
worldline. The black hole interior then becomes a baby universe
(rather than simply terminating at a siingularity), and
the exterior spacetime eventually settles back to the vacuum. A second
assumption is that quantum fluctuations of the geometry are small and
the notion of an approximate semiclassical geometry can be employed.
This  assumption follows  formally from a $1/N$ expansion, where $N$ is the
number of matter fields.
\vbox{{\centerline{\epsfxsize=3.3in \epsfbox{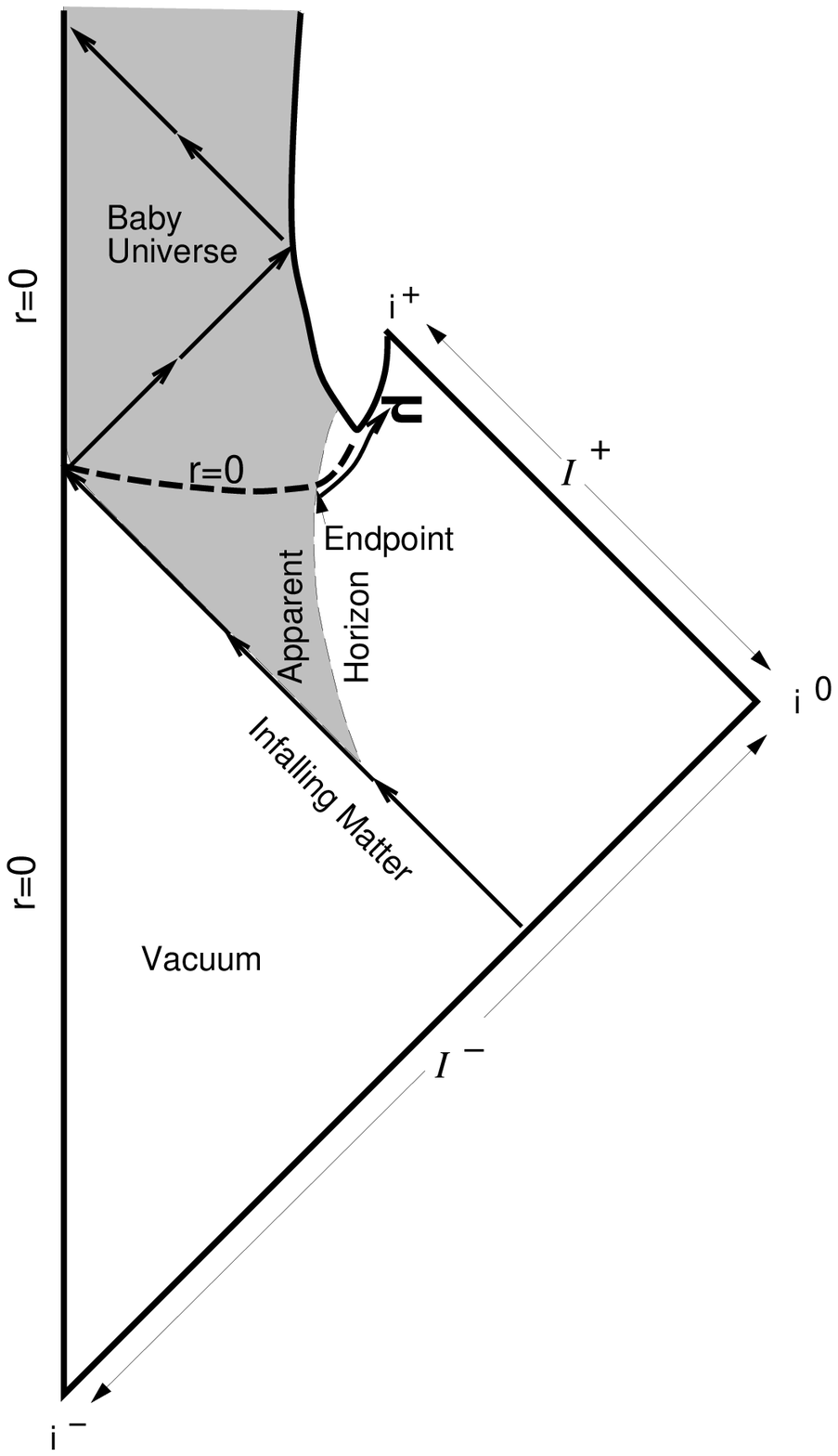}}}}
\begin{figure}[h]
\caption[]{A large infalling matter pulse forms a black hole (shaded
region) which evaporates down to zero size at the endpoint. Shortly
thereafter, the black hole interior splits off from the exterior
spacetime. The exterior spacetime settles back to the vacuum, and the
Bondi mass accordingly vanishes at $i^+$. $\tau$ measures the proper
time after the endpoint along the worldline indicated.}
\label{one}
\end{figure}

$g_s$ is a new parameter in the theory, which only affects
topology-changing processes. If (unnaturally) set
to zero, the umbilical cord can
never break, and information is stored within an eternal remnant.
We find this behavior implausible: in quantum mechanics what is not
forbidden is compulsory, and there is no conservation law which
forbids disassociation of the (neutral) black hole interior.

In such ($g_s\neq 0$) models there is an ``$S$-matrix'', denoted  $S$, which
maps the
incoming Hilbert space (on ${\cal I}^-$) to the tensor product of the
outgoing (on ${\cal I}^+$) and baby universe Hilbert
spaces\footnote{We assume here for convenience
that no baby universes are present initially.
A different  choice of initial state would affect the measure in equation
(6) below, but not our final conclusions.}. Important
subtleties arise in utilizing $S$ to describe scattering from ${\cal
I}^-$ to ${\cal I}^+$. As discussed in \cite{as}, there are at least
two inequivalent proposals.


The first proposal, advocated by Hawking, amounts to throwing away
whatever falls in to the black hole. For a single black hole, one
forms a $\$ $-matrix by simply tracing (denoted $tr_i$) over the
internal and unobservable baby universe Hilbert space:
\be
\$_1 =tr_i[{ S} { S}^\dagger].
\label{dllr}
\ee
$\$_1 $ acts on density matrices, and in general maps pure states to
mixed states. In general for an arbitrary number $n$ of
black holes, $\$ $ is defined by
\be
\$ =\sum_{n=0}^\infty tr_{i_1}tr_{i_2}\ldots tr_{i_n}[{ S} { S}^\dagger],
\label{ndllr}
\ee
with a separate trace for each internal Hilbert space.

This proposal has been criticized \cite{suss} on the grounds that it
will inevitably violate energy conservation. This criticism invoked
results of \cite{bps}. However \cite{bps} considered only
unitarity-violating dynamics which are strictly local in time. Black
hole formation and evaporation requires a finite time and so is not
local in this sense. If we try to derive a local description by
looking at time scales long compared to the formation/evaporation
time, the incoming states which create the black holes in the first
place are no longer present in the effective field theory. Thus ---
while we agree that energy conservation is an important issue here ---
we know of no regime in which the results of \cite{bps} are directly
applicable (although perhaps an adaptation of their arguments can be
applied). Hawking's proposal therefore remains a logical contender for
a consistent description of quantum black hole processes. It is of
course of utmost importance to determine whether or not this proposal
is fully consistent, but we shall not attempt to do so here.

In this paper we will develop an alternate proposal based on third
quantization \cite{kuch} of the baby universe Hilbert space, and
partially inspired by an analogy to string theory \cite{as}.  In this
formalism, baby universes are created and annihilated by
operators which act on the third-quantized Hilbert space. For the case
of a single black hole, this alternate proposal is indistinguishable
from Hawking's proposal. However for multiple black holes,
(\ref{ndllr}) is replaced by
\be
\$ =\sum_{\{n_k\}}\langle\{n_k\}|S|\{0\}\rangle
\langle\{0\}|S^\dagger|\{n_k\}\rangle = \sum_{n=0}^\infty
tr_{i_1}tr_{i_2}\ldots tr_{i_n} [\sum_{j=1}^{n!}P_j SS^\dagger],
\label{3dllr}
\ee
where $|\{n_k\}\rangle$ is the third-quantized state with
$n_k$ baby universes in the $k$th single-baby-universe state,
the operator $P_j$ generates the
$j$th permutation of the $n$ baby universes and the initial baby universe
state $|\{0\}\rangle$
is made explicit in the middle expression. These permutations arise
because the third-quantized baby universes are treated like
indistinguishable particles, unlike in (\ref{ndllr}) where they are
effectively treated as distinguishable. The difference between the
two proposals is schematically illustrated in figures 2 and 3.

\vbox{{\centerline{\epsfxsize=4.5in \epsfbox{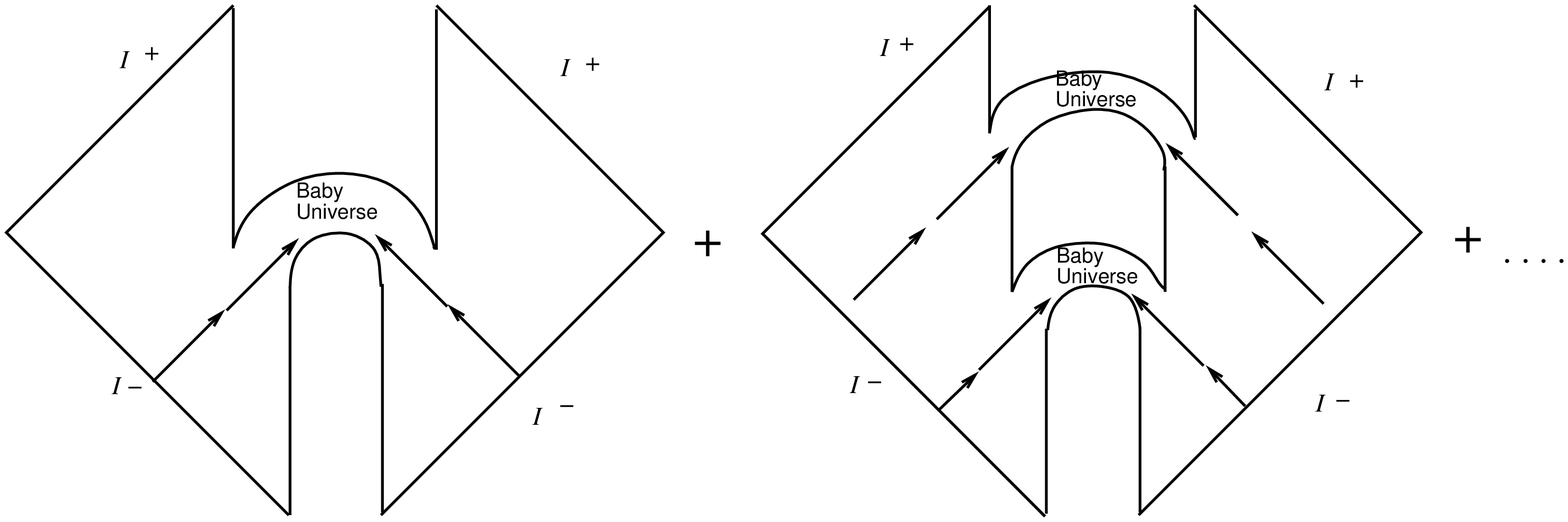}}}}
\begin{figure}[h]
\caption[]{In Hawkings proposal, a $\$$-matrix is formed by tracing over
everything which falls in to the black hole. This trace effectively
sews together the left and right portions (representing $S$ and $S^\dagger$)
of each diagram. Contributions to $\$$ arising from one or two black holes
are depicted.}
\end{figure}

\vbox{{\centerline{\epsfxsize=6.5in \epsfbox{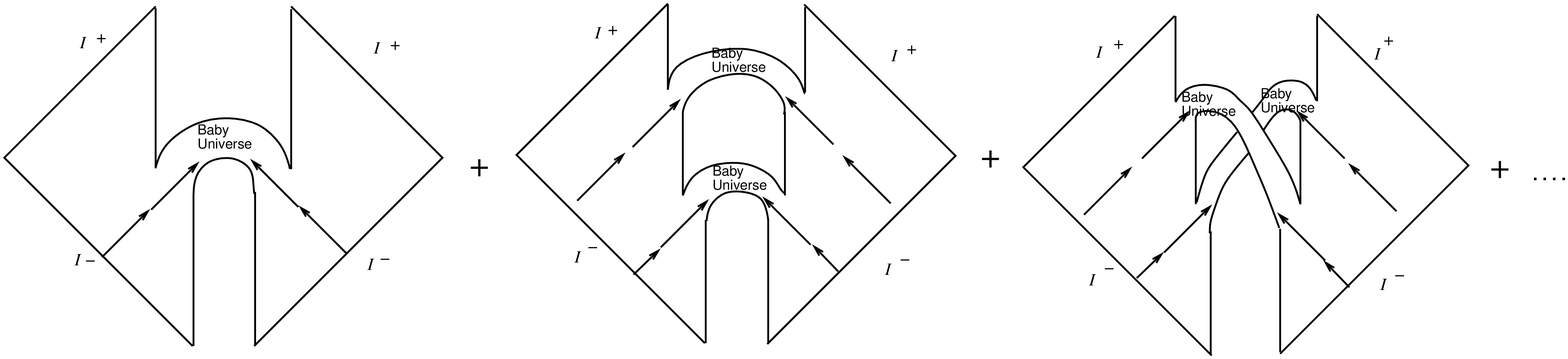}}}}
\begin{figure}[h]
\caption[]{Third quantization implies that the baby universes are
distinguishable only by their internal state, and not by the
spacetime location of the black hole from which they were created.
One accordingly must sew the left and right halves of the diagrams together in
all possible ways. This has no consequence for the one-black-hole
sector of $\$$, but for two black holes, there is one extra diagram,
as illustrated.}
\end{figure}

In reference \cite{suss}, expression (\ref{3dllr}) was criticized on
the grounds that the probabilities do not properly cluster: The last diagram
in figure 3 represents interactions between widely separated
experiments. This
lack of clustering accurately illustrates
the difficulties in  trying to use (\ref{3dllr}) to describe a theory with
information loss. However, the violations of clustering are
physically unobservable because the Hilbert space divides into $\$ $-matrix
superselection sectors, in each of which clustering is valid. To see this,
following \cite{worm}, let
$\phi_i$ denote the third-quantized operator which creates and
annihilates a single baby universe in the $i$th state. Consider an
``$\alpha$-basis'' $|\{\alpha\}\rangle \equiv | \alpha_1, \alpha_2,
\ldots\rangle$ for the baby universe sector of the third-quantized Hilbert
space
whose elements obey
\be
\phi_i |\{ \alpha\}\rangle = \alpha_i|\{ \alpha\}\rangle.
\label{abas}
\ee
In this basis, the ${ \$} $-matrix (\ref{3dllr}) becomes
\be
\$ = \int\prod_j d\alpha_j \langle\{\alpha\}|S|\{0\}\rangle
\langle\{0\}|S^\dagger|\{\alpha\}\rangle.
\label{wfk}
\ee
The interaction which describes the creation of a baby universe in the
$i$th state by a black hole is linear in the operator $\phi_i$. The
operator ${ S}$ therefore has vanishing matrix elements between
different $\alpha$-states. This allows us to write
\be
\$ = \int\prod_j \left( \frac{d\alpha_j}{\sqrt{2\pi}} e^{-\alpha_j^2/2}\right)
S_{\{\alpha\}}S_{\{\alpha\}}^\dagger,
\label{adol}
\ee
where
\be
\langle\{\alpha\}|S|\{\alpha'\}\rangle
\equiv \delta(\alpha - \alpha') S_{\{\alpha\}} .
\label{sdef}
\ee
The physical content of (\ref{adol}) is that the theory decomposes
into superselection sectors parameterized by  $\{ \alpha\}$, {\it
i.e.} the values of these parameters are not changed in any scattering
experiment. Furthermore, within each superselection sector, $\$ $
factorizes into the product of matrices ${ S}_{\{\alpha\}}$.

This observation has been made previously by many people following
\cite{worm} (perhaps in a slightly different form) and is not new to
the present work.  On its own this does not provide a resolution of
the information puzzle because it is far from obvious that the ${
S}_{\{\alpha\}}$ are unitary matrices. Indeed, previous estimates of
${ S}_{\{\alpha\}}$ (unpublished, or as obtained by the rules of
\cite{suss}) would seem to indicate that it does not even conserve
probability! In this case one would have to conclude that third
quantization simply makes no sense in the context of black hole
physics (no one promised us it would). However, in the following we
shall see that a careful evaluation of ${ S}_{\{\alpha\}}$ in the
type of models under consideration does in fact yield a unitary matrix. The key
feature (not considered previously) essential for unitarity is that
baby universe formation occurs with a finite quantum mechanical
amplitude per unit proper time, rather than instantaneously at the
black hole endpoint.

As a warm-up to computation of ${ S}_{\{\alpha\}}$, we first mention
some features of an initial massive particle $|I\rangle$ at rest which
decays to a number of possible final states $|F_i(t)\rangle$ of
outgoing particles with decay constants $g_i$. (Of course one can
always set all but one of the $g_i$s to zero by a basis rotation.) At
the moment $t_0$ at which the decay occurs, the outgoing particles are
created in some state $|F_i(0)\rangle$. At some later time $t$ they
will be in a different state (by virtue of their motion)
$|F_i(t-t_0)\rangle$. If the outgoing particles promptly disperse
after they are created, and the decay times are long compared to other
scales in the problem, we may make the approximation that states at
different times are orthogonal:
\be
\langle F_i(t')|F_j(t)\rangle = \delta_{ij}\delta(t'-t).
\label{fnrm}
\ee
In this same approximation the outgoing and initial state do not
interact after the decay has occurred, and the interaction Hamiltonian
is characterized by the matrix elements
\be
\langle I|H_{int}|F_j(t)\rangle = ig_j\delta(t).
\label{hinr}
\ee
Solving Schr\"odinger's equation we then find that the full quantum
state is\footnote
{The result (\ref{psfn}) is equivalent to Fermi's golden rule, but adapted
to the situation that the system has a semiclassical motion parameterized by
$t$
   .
Defining energy eigenstates $|E,i\rangle = \int_{-\infty}^\infty dt\,e^{iEt}
F_i(t)$, the golden rule decay rate
$\Gamma = 2\pi \rho(E_I) \sum_i |\langle E,i| H_{int} | I \rangle|^2$,
where $\rho(E_I) = 1/2\pi$, is the same as in (\ref{psfn}).}
    ,
\be
|\psi(t)\rangle = a(t)|I\rangle - \int_0^t dt_0 a(t_0)
\sum_j g_j |F_j(t-t_0)\rangle ,\quad a(t)=e^{-\sum_i g_i^2 t/2}.
\label{psfn}
\ee

The decay of a black hole spacetime to an exterior spacetime plus a
baby universe is similar, except that one of the decay products is
emitted into an $\alpha$-state (rather than the vacuum)
and there is an additional suppression
arising from overlap of the initial and final state wave functions. A
natural time parameter is the proper time $\tau$ from the black hole
endpoint along the worldline of the umbilical cord (see figure 1).
After the umbilical cord breaks, $\tau$ is chosen to be the proper
time at the (newly-formed) origin. (It will not be necessary to choose
a specific time parameter prior to  the endpoint.) We consider a
Hamiltonian which evolves the system along a sequence of
asymptotically flat spacelike slices labeled by the value of $\tau$ at
which the slice intersects the umbilical cord or the origin. For
$g_s=0$ (so that decay is suppressed) the quantum state
$|I(\tau)\rangle$ on a slice at time $\tau$ can be written as a state
in the tensor product of the Hilbert space inside and outside the
umbilical cord:
\be
|I(\tau)\rangle = \sum_{m,J}\rho_{mJ,I}(\tau)|m\rangle |J\rangle,
\label{tprod}
\ee
where $m$ ($J$) is an index in the  internal (external) Hilbert space.

We wish to evolve the full state $|\psi(\tau)\rangle$ for $g_s \neq 0$
(so that decay can occur) in a baby universe $\alpha$-state.
The
(third-quantized)
interaction Hamiltonian $H_{int}$
contains a piece which destroys the incoming state from ${\cal I}^-$
and creates a baby universe and a state outgoing to ${\cal I}^+$. It is
accordingly linear
in the the baby universe field operator $\phi_i$. In an
$\alpha$-state, $\phi_i$ can be replaced by its eigenvalue
$\alpha_i$. With this replacement, the interaction Hamiltonian
describes the decay of the incoming state $|I(\tau)\rangle$ to an
outgoing state in the exterior spacetime:
\be
\langle J(\tau')|H_{int}|I(\tau)\rangle = ig_s\sum_i
\alpha_i\rho_{iJ,I}(\tau)\delta(\tau' ).
\label{mlmnt}
\ee
$|J(\tau')\rangle$ here is the unitary evolution of the detached
exterior state $|J\rangle$ (using the Hamiltonian which
incorporates the appropriate reflecting boundary conditions
\cite{verdan,sthor} at the newly-formed origin) for a time
$\tau'$ after the decay has occurred, which approximately
(at large $N$)
obeys $\langle J(\tau')|J(\tau)\rangle=\delta(\tau'-\tau)$ as in
(\ref{fnrm}). The
$i$ index here runs over a smaller set of values than the
corresponding $m$ index in (\ref{tprod}) because it only includes
states obeying the appropriate  boundary conditions at the umbilical
cord. This distinction will
be further discussed below.

The full quantum state is determined from the
Schr\"odinger-Wheeler-DeWitt equation to be
\be
|\psi(\tau)\rangle = |\psi_1(\tau)\rangle + |\psi_2(\tau)\rangle,
\label{fnll}
\ee
where
\beq
|\psi_1(\tau)\rangle &=& \sum_{I'} |I'(\tau)\rangle a_{I'I}(\tau) \nonumber\\
|\psi_2(\tau)\rangle &=& - g_s\int^\tau_0 d\tau_0
\sum_{i,J,I'}\alpha_i\rho_{iJ,I'}
|J(\tau-\tau_0)\rangle a_{I'I}(\tau_0),
\eeq
with $a_{I'I}(\tau)$ the time-ordered exponential
\be
a(\tau) = T e^{-g_s^2 \int_0^\tau d\tau'\, \Gamma(\tau')/2}, \qquad
\Gamma_{I'I}(\tau) = \sum_{J,i,i'}
\alpha_{i'} \alpha_{i} \rho^*_{i'J,I'}(\tau) \rho_{iJ,I}(\tau)
   .
\ee
By construction $\langle\psi(\tau)|\psi(\tau)\rangle =1$, so that expression
(\ref{fnll}) provides a unitary
description of black hole formation/evaporation.

Note that the $\alpha$-parameters are not directly equal to in-out
$S$-matrix elements, but rather enter them in a complicated and
indirect fashion in (\ref{fnll}) through the decay constants. It
should now be evident that a quantum-mechanically variable decay time
is crucial for unitarity of the $S_{\{\alpha\}}$. For example in the
$RST$ model \cite{rst}, where decay occurs instantaneously at the
evaporation endpoint, the $S_{\{\alpha\}}$ defined in (\ref{sdef})
would not be unitary.

The probability that the interior eventually splits off in to a baby
universe is $1-a_{II}^2(\infty)$. It is important that this is unity, in
order to avoid an eternal remnant with probability one.
It appears from (\ref{fnll}) that this will indeed be the case for
generic values of the $\alpha$-parameters. However it
is worth noting that counterexamples are easily constructed if the
$\alpha$-parameters respect global symmetries. For example suppose
that the $\alpha$-parameters were flavor-blind and took the same
values for a black hole formed by a collapsing
$\ket{\mbox{\em{chocolate}}}$ state and a collapsing
$\ket{\mbox{\em{vanilla}}}$ state. Then they will all vanish for the
collapsing state $(\ket{\mbox{\em{chocolate}}} -
\ket{\mbox{\em{vanilla}}})/{\sqrt 2}$, which accordingly never decays.
Consistency of our picture thus requires a genericity condition on the
$\alpha$-parameters.

In practice, the $\alpha$-parameters are not known initially. If only
a small number (relative to the number of relevant
$\alpha$-parameters) of experiments are performed, the results
predicted by our formulae are indistinguishable from those of
Hawking's. Differences will emerge only when the number of experiments
is of order the number of relevant $\alpha$-parameters. In fact since
there are an infinite number
they can never all be measured.
However,
``most'' of the $\alpha$-parameters have a very small effect on the
outgoing quantum state because the incoming state has a very
small amplitude for producing the corresponding baby universe.
These parameters will be very hard to measure but, by the
same token, they will have little effect on the out-state. We expect
that if one repeatedly prepares identical collapsing states, the
outcome will be increasingly predictable. This is the  case in any real
experiment: the outcome is affected by an infinite number of higher-dimension
operators whose coefficients we do not know,
but which have little effect on the outcome. However a precise
understanding of how the predictability increases is lacking at
present.



It would be of great interest to obtain a measure of the number of
$\alpha$-parameters relevant for the prediction of the out-state
associated with a given in-state.  An upper bound on the
number can be estimated as follows.  The information arrives at
${\cal I}^+$ in the decay time $\tau_D$, using radiation with
total energy and angular momentum of order one (in powers of $M$).
Standard thermodynamic estimates imply that there are of order
$e^{\sqrt {N \tau_D}}$ such states.
So the number
of relevant parameters is at most $e^{\sqrt{ N \tau_D}}$.
The number
of baby universe states may be greater than this, but only this finite number
of linear combinations of the $\alpha_i$ is relevant.  The number of baby
universe states depends on the detailed dynamics.  In the model of \cite{as},
the baby universe continues to expand after the endpoint forms, so the number
of
states is comparable to the maximum above.  With different dynamics, such that
the baby universe did not continue to expand, the number would be much smaller
and fewer experiments would be needed.

The notion that information is not really lost in black hole
formation/evaporation has been previously advocated by a number of
authors \cite{page,thoo,verl,stu}. In these works it was argued that
precise knowledge of the local laws of physics ({\it e.g.} string
theory) would eventually enable one to unitarily predict the out-state
from the in-state. In our proposal, this is not possible. Additional
input --- namely the values of the $\alpha$-parameters --- is
required\footnote{Although in the proposal of \cite{stu} these might be
viewed as non-perturbative parameters of string theory.}. A further
distinction is that in these previous works the information comes back out
before the endpoint, whereas in our picture (as we shall see) it comes out
after the endpoint. Thus there is no obvious connection of our results
with these previous works. Nevertheless it is possible that future
work will reveal a unified treatment of these different pictures.

The possibility of baby universe formation has no effect on the
quantum state $|\psi\rangle$ in (\ref{fnll}) prior to the future light
cone of the black hole endpoint, since formation  cannot occur prior
to this point. The entropy on ${\cal I}^+$ prior to the future of the
endpoint
accordingly
is independent of $g_s$ and the $\alpha$-parameters. It follows (as
expected from causality) that the information contained in the
collapsing state can not have been returned to ${\cal I}^+$ prior to
the future of the endpoint. The rate at which it can come out
after the endpoint encoded in the finite amount of available
energy is highly constrained by entropy/energy bounds \cite{cash}.
Consistency with these bounds
then implies that all models of the type we discuss have long-lived
remnants. This is not obvious from equation (\ref{eq14}). However in two
dimensions it can be seen explicitly, as follows.

Before the decay, the quantum state of the matter fields is in the
Hilbert space ${\cal H}$ of states on the half-line extending from the
origin. Momentarily after the decay, it is in the product space ${\cal
H}'$ of baby universe and exterior Hilbert spaces, which obey
{\it e.g.} Neumann
boundary conditions along the umbilical cord. ${\cal H}'$ may be
regarded as a subspace of ${\cal H}$, and the quantum state prior to
the decay is a general state in ${\cal H}$ which has a component in
${\cal H}'$. Decay can occur only through this component.
Denoting the projection from
${\cal H}$ to
${\cal H}'$ by ${\cal P}$, the decay
rate will accordingly contain a factor $\matel{I}{{\cal P}}{I}$.
More explicitly, the $\alpha$-ensemble average of the decay rate is
\be
\int\prod_j \left( \frac{d\alpha_j}{\sqrt{2\pi}} e^{-\alpha_j^2/2}\right) g_s^2
\sum_J \Bigl|
\sum_i \alpha_i\rho_{iJ,I}\Bigr|^2 = g_s^2
\matel{I}{{\cal P}}{I}
\label{eq14}
\ee
The projection $\matel{I}{{\cal P}}{I}$ can be represented as a
Euclidean path integral
with initial and final boundary conditions corresponding to the state
$\ket{I}$. The intermediate projection is represented by the insertion
of a circular puncture at the boundary of which Neumann boundary
conditions are imposed on the matter fields, together with appropriate
boundary conditions on the gravitational fields. To compute this path
integral we must choose a coordinate system. The simplest choice is
``sigma coordinates'', in which the matter vacuum is simply the state
annihilated by positive $\sigma^\pm$ Fourier components of the matter
fields. The metric is $ds^2 =
-e^{2\rho_\sigma}d\sigma^+d\sigma^-$, and $\rho_\sigma$ goes to zero
on ${\cal I}^-$. The path integral is then proportional to the
determinant of the matter laplacian regulated with respect to
$\rho_\sigma$:
\be
\matel{I}{{\cal P}}{I} \sim \det[\Box]_{\rho=\rho_\sigma}.
\label{eq15}
\ee
The formula for the trace anomaly on flat manifolds with curved
boundaries then implies
\be
\det[\Box]_{\rho=\rho_\sigma} \sim e^{-N\oint K\rho_\sigma/12\pi}
\det[\Box]_{\rho=0},
\label{eq16}
\ee
where the exponent contains the integral of the extrinsic curvature
$K$ around the boundary of the puncture, and $N$ is the central charge
of the matter fields. The determinant evaluated at $\rho=0$ is
independent of the black hole mass to leading order and does not
concern us. For a small puncture, $\rho$ is nearly constant along the
boundary, and we may approximate
\be
\matel{I}{{\cal P}}{I} \sim e^{-N\rho(\sigma_0)/6}.
\label{eq17}
\ee
where  $\sigma_0$ is  the  location of the puncture\footnote{This calculation
hides a divergent, cutoff-dependent factor which is absorbed by multiplicative
renormalization of $g_s$.}. Explicit
computation \cite{rst,fpst} reveals that near the evaporation endpoint
\begin{equation}
N\rho_\sigma(\sigma_0)/6 \sim N M/3  \sim S_{tot},
\end{equation}
where $S_{tot}$ here is the fine-grained ``entropy of entanglement''
($-tr\rho_{out} \ln \rho_{out}$)
of the quantum state $\rho_{out}$ outside the black hole. $\rho_{out}$
is obtained by tracing over the portion of the quantum state inside the
black hole, and $S_{tot}$ is the total entropy in outgoing Hawking radiation.
The exponentially large (in $M$) value of $e^\rho$ near the endpoint
is equivalent to the well-known fact that (in any dimension)
frequencies of field modes undergo exponentially large redshifts
in propagation from ${\cal I}^-$ to the vicinity of the endpoint.
So we see that
\begin{equation}
\matel{I}{{\cal P}}{I} \sim e^{-S_{tot}} <<1,
\label{psen}
\end{equation}
and the decay time $\sim e^{S_{tot}}/g_s^2$
is extremely long.\footnote{To those familiar with
string theory, the preceding discussion is just the usual statement
that the string coupling is field dependent.}
\footnote{It is possible that the suppression of the decay rate can alternately
   be
understood as a consequence of the need to conserve energy, and in this way
is related to results of \cite{rmnt}.  In order to carry away the infalling
information, a large number of low energy outgoing particles are needed.
A phase space suppression might be then be expected, and this would make its
appearance in the matter determinant. We further note that information flow
in and out of the black hole is accompanied by energy flow, in harmony
with remnant constraints discussed in \cite{rmnt}.}

We stress that what we have computed is the decay time in the rest frame
of the umbilical cord. In general there will be a dilation factor relating
this to the decay time as seen at ${\cal I}^+$. The precise form of this
factor appears rather model-dependent, but we do not expect it to affect
the exponentially growing behavior.
Compatability with the information/energy bounds \cite{cash} only requires
that the decay time as seen at ${\cal I}^+$ must grow as
a (dimension dependent) power of $S_{tot}$.

Relation (\ref{psen}) states that the entanglement of the interior
and exterior of the black hole slows down the decay rate.
A heuristic understanding of this can be obtained
without resorting to explicit calculation, as follows.
Divide the matter field modes into those which are fully inside or outside
the apparent horizon of the black hole and those which  overlap
the horizon. As discussed  in detail in \cite{fpst}, the entropy obtains
contributions only from the overlapping modes (which contribute to
the entanglement) and can be written as a sum with equal
contributions from each mode. The sum diverges and an ultraviolet
regulator is needed to define it. The regulator dependence can be
absorbed by a shift in the zero of the entropy. The physically relevant
finite part of the entanglement entropy increases as the black hole
evaporates because the increasing redshift of  field modes at the horizon
relative to ${\cal I}^-$ causes increasing numbers of them to
contribute  to the entropy.

Similarly, ${\cal P}$ is a product of projection operators which are
unity except for the overlapping modes. $ln \matel{I}{{\cal P}}{I}$  is then a
sum over modes which  obtains
contributions only from the overlapping modes. At very high frequencies
there should be  equal
contributions from each mode. One thus expects that $ln \matel{I}{{\cal P}}{I}$
is proportional to $S_{tot}$, as we have verified by explicit calculation.

The heuristic arguments of the preceding two paragraphs
are still applicable in $3+1$
dimensions with slight modifications (and
conceivably might be crystallized in to
a precise calculation). Thus we expect that
the decay time is still of order $e^{S_{tot}}$, where the fine grained entropy
produced in
$3+1$ dimensions is $S_{tot}=16\pi M^2 /3$. This greatly
exceeds the information
bound of $M^4$ found in \cite{cash}.

Two main objections have been raised in the past to proposed resolutions
of the information puzzle which involve remnants. The first is the
problem of huge pair production rates or virtual loop effects
associated with the large
numbers of long-lived states. Although this issue deserves further
scrutiny in the present context, it appears plausible to us that these effects
will
be suppressed by a mechanism of the type described in \cite{bos,rmnt} :
Roughly speaking, the large
number of states are in a distant region deep inside the black hole
and most of them cannot be accessed (by causality)
in any finite time process.

The second objection has been the lack of a good dynamical reason why
a long-lived remnant
would stay around long enough to reemit the information. Naively,
the cost in action for a (neutral) Planckian remnant to disappear is of order
one and it should therefore  disappear in a time of order the Planck time.
In this paper we have not only found a general dynamical origin of
the long decay time, we have also described the actual mechanism
by which the information is reemitted.

In conclusion, third quantization appears to offer a viable resolution
to the black hole information puzzle. We find it fascinating that the
consistency of quantum mechanics and gravity in our own universe may
require the existence of other universes.

\subsection*{Acknowledgments}

This work was supported in part by DOE grant DOE-91ER40618 and NSF
grants PHY91-16964 and PHY89-04035. A.S. would like to
acknowledge the hospitality of the Newton Institute where part of
this work was
completed. We are grateful to T. Banks, S. Giddings, S. Hawking, D. Lowe,
J. Preskill, M.
Srednicki, L. Susskind and L.~Thorlacius for useful discussions.

\gdef\journal#1, #2, #3, 1#4#5#6{       
 {\sl #1~}{\bf #2}, (1#4#5#6) #3}       


\begin{thebibliography}{}
\bibitem{hawk}S.W.~Hawking, \journal Comm. Math. Phys., 43, 199, 1975.

\bibitem{hawktwo}S.W.~Hawking,\journal Phys. Rev., D14, 2460, 1976 .

\bibitem{worm} S.~Coleman, \journal Nucl.~Phys., B307, 864, 1988;
 S.B.~Giddings and A.~Strominger, \journal Nucl.~Phys., B307, 854, 1988.

\bibitem{cash} Y. Aharanov, A. Casher and S. Nussinov, \journal
Phys. Lett. , 191B, 51, 1987; R. D. Carlitz and R. S. Willey, \journal
Phys.~Rev., D36, 2336, 1987.

\bibitem{as} A.~Strominger, unpublished.

\bibitem{suss} L.~Susskind, {\it Comment on a Proposal By Strominger}, preprint
   SU-ITP-94-14, hep-th/9405103.

\bibitem{bps} T.~Banks, M.~Peskin, and L.~Susskind, \journal
Nucl.~Phys., B244, 125, 1984.


\bibitem{kuch} K.~Kuchar, \journal J.~Math.Phys., 22, 2640, 1981;
N.~Cadderni and M.~Martellini, \journal
Int.~J.~Theor.~Phys., 23, 23, 1984;  A. Jevicki, in {\it Frontiers in
Particle Physics '83},
D. Sijacki et. al., eds., World Scientific (1984);
I. Moss, in {\it Field Theory, Quantum Gravity and Strings, II}
eds. H. deVega and N. Sanchez,
Springer, Berlin (1987); T.~Banks, \journal Nucl.~Phys.,
B309, 643, 1988; V. Rubakov, \journal Phys. Lett., B214, 503, 1988;
S.B.~Giddings and A.~Strominger, \journal Nucl.~Phys.,
B321, 481, 1989; W. Fischler, I. Klebanov, J. Polchinski and L. Susskind
\journal Nucl.~Phys.,
B327, 157, 1989; A.~Strominger, \journal Phil.~Trans.~R.~Soc.~Lond.,
A329, 395, 1989.

\bibitem{verdan}T.~D.~Chung and H.~Verlinde,
{\it Dynamical Moving Mirrors and Black holes},
preprint PUPT-1430,
hepth/9311007.

\bibitem{sthor} A.~Strominger and L.~Thorlacius, {\it Conformally Invariant
Boundary Conditions for Dilaton Gravity} ITP preprint
NSF-ITP-94-34,
hep-th/9405084.

\bibitem{rst} J.G.~Russo, L.~Susskind, and L.~Thorlacius, \journal
Phys.~Rev., D46, 3444, 1993; \journal Phys.~Rev., D47, 533, 1993.

\bibitem{page} D.~Page, \journal Phys.~Rev.~Lett, 44, 301, 1980.

\bibitem{thoo} G.~'t Hooft, \journal Nucl.~Phys., B335, 138, 1990.

\bibitem{verl} E. Verlinde and H.
Verlinde, \journal Nucl.~Phys., B406, 43, 1993.

\bibitem{stu}  L.~Susskind,  L.~Thorlacius and J.~Uglum,
\journal Phys.~Rev., D48, 3743, 1993.

\bibitem{fpst} T.~Fiola, J.~Preskill, A.~Strominger, and S.~Trivedi,
{\it Black Hole Thermodynamics and Information Loss in Two Dimensions}
preprint CALT-68-1918, hep-th/9403137, to appear {\it Phys. Rev.} {\bf D}.

\bibitem{rmnt} S. B. Giddings,
\journal Phys. Rev., D49, 4078, 1994 and private communication.

\bibitem{bos} T. Banks, M. O'Loughlin and A.~Strominger,
\journal Phys. Rev., D47, 4476, 1993.


\end{thebibliography}
\end{document}